# Non-interferometric accurate phase imaging via linear-convergence iterative optimization

JIANHUI HUANG,[1, 2] AN PAN,[3] HUILIANG JIN,[1] GUOXIANG MENG,[1] AND QIAN YE[1,*]

[1]School of Mechanical Engineering, Shanghai Jiao Tong University, Shanghai, China, 200240

[2]Optics Laboratory, Georgia Institute of Technology, School of Electrical and Computer Engineering, 777 Atlantic Drive N.W., Atlanta, Georgia 30332-0250, USA

[3]Pioneering Interdiscipline Center (PIC), State Key Laboratory of Transient Optics and Photonics, Xi'an Institute of Optics and Precision Mechanics, Chinese Academy of Sciences, Xi'an 710119, China

*yeqian@sjtu.edu.cn

## Abstract

This paper reported a general non-interferometric high-accuracy quantitative phase imaging (QPI) method for arbitrary complex-valued objects. Given by a typical 4-*f* optical configuration as the imaging system, three frames of small-window phase modulation are applied on the object's Fourier spectrum so that redistributed intensity patterns are produced on the image plane, in which the object phase emerges at different degree. Then, an algebraic relationship that connects the object phase with the output intensity is established to provide us with an approximate closed-form phase recovery. Further, an efficient iterative optimization strategy is developed to turn that approximate solution into an accurate one. Due to the linear convergence property of the iteration, a high-accuracy phase recovery is achieved without requiring heavy iterations. The feasibility and accuracy of the proposed method are verified by both numerical simulations and experiments on diverse phase objects.

## 1. Introduction

Quantitative phase imaging (QPI) has a wide range of applications biomedical imaging, which has been attracting increasing interests and making continuous progress [1-4]. It is a promising tool in label-free and stain-free cell imaging [3, 5], three-dimensional tomographic imaging [6-10], neuroimaging [11], etc. In these fields, the structure information of the tested sample is carried mainly by the phase, while the phase cannot be measured directly by current technology. The most common solution is to recover the hidden phase information computationally from the coded intensity patterns, where the recovery algorithm is dependent on a certain optical configuration and the corresponding relationship (decoding method) between the unknown phase and the measurable intensity. The object phase can be extracted from the measured intensity by many means. If there is an external and stable coherent reference wave, interferometry can be applied as a very convenient and accurate QPI technique, i.e., digital holography. This classical concept has been developed maturely into various practical schemes, such as off-axis digital holography [12], inline holography [13-15], phase-shifting holography [16, 17], wavefront-divisional digital holography [18, 19]. Specially, holographic QPI can be used in situations where there is no external reference wave. In these cases, such as diffraction phase microscopy [20, 21] and point-diffraction interferometry [22, 23], the reference wave is generated from zero-order diffracted wave of the object itself.

Alternatively, non-interferometric phase imaging is developed to be applied in the case lack of reference wave [24-26]. As the first non-interferometric phase imaging technique, Zernike phase contrast (ZPC) pioneered visual observation to transparent samples by applying $\pi/2$ phase shifting on the object's Fourier spectrum [27]. The phase-shifting occurs on the tiny Airy spot so that only the background wave is modulated while the diffracted wave stays unaffected. As a result, the object phase appears on the intensity image in a near-linear way. Due to the first-order Taylor approximation made on the phase, however, ZPC provides us only a qualitative phase recovery and is feasible to weak objects only. Such severe drawback limits further applications of ZPC in modern biomedical imaging. In fact, due to the lack of direct reference wave, non-interferometric QPI needs to measure more redistributed intensity patterns to provide sufficient constraints to guarantee the unique phase recovery solution. The additional intensity is often measured by modulating the object or introducing defocus. By applying more phase-shifting modulations, the original qualitative phase contrast imaging can be improved into a quantitative one that is available for detecting phase object with large phase range. This concept can be found in generalized phase contrast [24, 28], phase-shifting Zernike phase contrast [29, 30] and spatial light interference microscopy [31]. Besides, weak-defocus is also an effective QPI strategy as it establishes a deterministic equation that connects the object phase and the diffracted intensity. The transport of intensity equation (TIE) based QPI is a representative technique that obtains analytical phase recovery by using weak-defocus measurement [32, 33]. The improved TIE-QPI even allows single-shot analytical phase recovery based on the application of white light [34], weak-diffuser [35, 36] and beam splitting [25]. Note that the weak-defocus scheme is actually a special case of phase modulation on the object. TIE-QPI implemented by the means of paraboloid phase modulation simplifies and accelerates the measurement process significantly [37]. There are some difficulties in acquiring accurate phase recovery via non-interferometric QPI. For example, in the phase-shifting phase contrast based QPI, the reference wave is produced by modulating the object's Fourier spectrum within a small region, known as zero-order diffracted wave. It is close but not completely equal to an ideal planar or sphere wave. Obviously, the wavefront distribution of this kind of reference wave relates to the tested object, where different objects generate a reference wave with different distributions on both amplitude and wavefront. So there will be inevitable error in phase recovery if deals with that zero-order diffracted wave as an independent good reference wave. Similarly, the boundary condition and difference approximation involved in TIE-QPI act as a disadvantage to achieve accurate phase recovery. Provided enough object modulations and corresponding measured diffracted intensities, it is surely that the object phase can be solved by iterative phase retrieval. However, iterative phase retrieval suffers significant drawbacks such as initial sensitivity and stagnation, which makes it hard to achieve accurate phase recovery in few time [38].

This study proposes a non-interferometric high-accuracy phase-shifting QPI method by introducing iterative phase retrieval to improve the approximate phase recovery obtained by algebraic phase retrieval [26]. At first, a direct closed-form phase recovery is solved through algebraic phase retrieval. By establishing a rigorous phase-intensity equation between the object phase and the measured output intensities, the algebraic phase retrieval allows us reconstruct the complex object only by algebraic calculation. To improve accuracy, the phase-shifting area is set smaller than the object's Airy spot that is usually used in typical phase contrast imaging techniques. In this way, the rough algebraic phase recovery is sufficiently accurate to be used as an excellent initial phase guess of iterative phase retrieval. Therefore, accurate QPI may be obtained if we apply iterative optimization to the algebraic phase recovery. Based on this idea, a high-accuracy QPI method is finally developed by combining algebraic phase retrieval and iterative optimization. Taking the classical 4-$f$ phase contrast imaging system shown in Fig. 1 as the example, phase shifting is performed on the small-window area within the Airy spot of the object's Fourier spectrum for three times, and three intensity patterns are measured correspondingly on the output plane, each of which reflects the phase at different degrees. The flat output diffracted wave generated by modulating the object's spectrum within a small area serves as a good reference wave. Based on intrinsic trigonometric identity formed by the measured intensities, an analytical solution of the reference amplitude is derived, which is more accurate and general compared to generalized phase contrast method [28] and the illumination format now is out of consideration. Then a high-efficiency linear-convergence iterative optimization is designed to improve the accuracy of the reference wave continuously, which guarantees a high-accuracy quantitative phase recovery meanwhile.

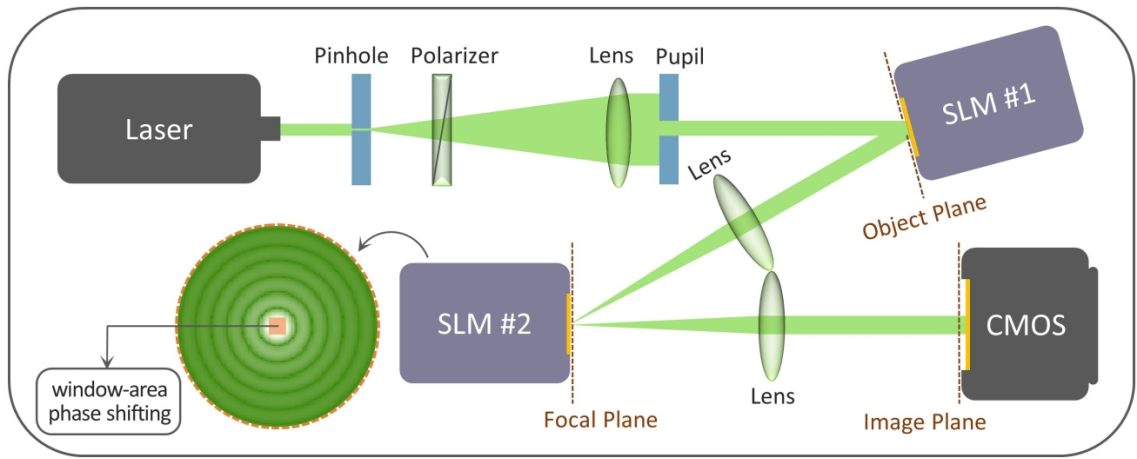

Fig. 1. Theoretical and experimental schematic diagram of the proposed non-interferometric high-accuracy QPI method. The window-area for phase-shifting should be small enough. A commercial laser of 532-nm wavelength plays as the coherent light source. Two SLMs (labelled as SLM #1 and SLM #2) are applied to load phase object and phase modulation, respectively.

## 2. Method

### *2.1 Problem setup and definition*

As illustrated in Fig. 1, this study discusses a kind of reference-free exact phase imaging method. It involves four devices, including a laser source, two SLMs and a CMOS image sensor. The raw laser beam is filtered by a pinhole to be smooth in amplitude and aberration-less in phase. Then it is collimated by a lens into a parallel beam, and finally shaped by a pupil into a circle coherent beam. On the object plane, a 2D phase object is loaded on SLM #1 in the format of a grayscale image. Another SLM, labelled as SLM #2, is placed on the focal plane and plays the role of shifting the phase of the object's Fourier spectrum within a small window. After modulation, a CMOS image sensor is arranged to record the intensity distribution.

Generally, the phase object is defined by a complex-valued function as $Ce^{iE}$, where $C$ is the amplitude, and $E$ the phase. The Fourier spectrum of the object is defined by $he^{i\alpha}$. The phase modulation is defined by $e^{itw}$, where $w$ indicates the phase-shifting window, and $t$ is the phase-shifting value in radian. The optical field on the image plane is defined by $Ue^{iV}$, whose intensity is defined by $I$, $I = U^2$. It is assumed there is no phase aberration on the filtered illumination laser beam, so the incident beam is expressed as a real function $B$. In this paper, the functions on the object plane and image plane are defined on space domain $(x, y)$, and the functions on the focal plane are defined on spatial frequency domain $(k_x, k_y)$. For a clear distinction, functions in spatial domain are written in capital, while functions in spatial frequency domain are written in lowercase. Besides, the coordinates are omitted for simplicity. It is reasonable to regard the background amplitude and the object amplitude as a combination, namely, $\overline{C} = BC$. Suppose the lens acts as an ideal Fourier transform, the intensity image captured by the CMOS is modelled as

$$Ue^{iV} = \mathsf{F}\left(\mathsf{F}\left(\overline{C}e^{iE}\right) \cdot e^{itw}\right) = \mathsf{F}\left(he^{i\alpha} \cdot e^{itw}\right), \tag{1}$$

where the symbols $\mathcal{F}$ refers to 2D Fourier transform. Expand the phase modulation $e^{itw}$ as $e^{itw} = 1 - w + e^{it}w$, and hence Eq. (1) is calculated as $Ue^{iV} = \overline{C}e^{iE} - Ke^{iP} + e^{it}Ke^{iP}$. Here the complex-valued function $Ke^{iP}$ is defined as the Fourier transform of $he^{i\alpha}w$, i.e., $Ke^{iP} = \mathcal{F}(he^{i\alpha}w)$, which indicates the output field when blocks the focal plane spectrum with the window $w$. Calculate the intensity on both sides and we obtain

$$I(t) = \overline{C}^2 + (2\cos t - 2)\left[\overline{C}K\cos(E-P) - K^2\right] + (2\sin t)\overline{C}K\sin(E-P). \tag{2}$$

This equation establishes the algebraic relationship between the output intensity pattern $I(t)$ and the phase-shifting value $t$. For a particular case $t = 0$, it means no phase modulation is applied on the object's Fourier spectrum so that the measured intensity pattern is precisely the square of the object's amplitude. It is very hard to recover the object phase $E$ through only one intensity measurement $I$. A widely adopted strategy is taking more phase modulations so that forms more constraints. In this paper, we present a kind of non-interferometric high-accuracy QPI based on multi-value small-window phase shifting on the object's Fourier spectrum. The concept of introducing a window-shape phase-modulated intensity measurement to help solve the phase map of a complex-valued object is firstly proposed in [39], called phase diversity. Here the use of small-window phase modulation is the condition of algebraic phase retrieval, by which a rigorous intensity-phase relationship is established to solve a quantitative phase recovery.

### *2.2 Algebraic phase retrieval*

According to the phase-intensity relationship Eq. (2), three phase-shifts with different values of $t$ is adequate to form a determined linear equation with respect to the unknown quantities, i.e.,

$$\begin{bmatrix} I_1 \\ I_2 \\ I_3 \end{bmatrix} = \begin{pmatrix} 1 & 2\cos t_1 - 2 & 2\sin t_1 \\ 1 & 2\cos t_2 - 2 & 2\sin t_2 \\ 1 & 2\cos t_3 - 2 & 2\sin t_3 \end{pmatrix} \begin{bmatrix} \overline{C}^2 \\ \overline{C}K\cos(E-P) - K^2 \\ \overline{C}K\sin(E-P) \end{bmatrix} = T \begin{bmatrix} R_1 \\ R_2 \\ R_3 \end{bmatrix}. \tag{3}$$

In this relationship, the output intensities $\{I_1, I_2, I_3\}$ and the phase-shifting values $\{t_1, t_2, t_3\}$ are known, while the quantities on the right column vector, $\overline{C}$, $K$, $E$ and $P$, are remaining to be solved. Record the coefficient matrix as $T$ and express the unknown quantities as $\{R_1, R_2, R_3\}$. It is proved that $T$ is always invertible as long as the inputting three $t$ are different from each other. Therefore, $R$ can be determined uniquely through matrix inversion calculation to $I$, $R = T^{-1}I$. Note that the object amplitude $C$ can be recovered directly from $R_1$, $C = \sqrt{R_1}/B$. And the solution for the object phase $E$ should be

$$E = P + \arctan \frac{R_3}{R_2 + K^2}. \tag{4}$$

Note that here the 'arctan' calculation is 'vector-arctan' that gives a result ranging from $-\pi$ to $\pi$ instead of $[-\pi/2, \pi/2]$ of typical 'arctan'. To reconstruct the phase, we should obtain the solution of $P$ and $K$ at first. According to the trigonometric identity $\sin^2 + \cos^2 \equiv 1$, an indefinite solution to $K$ (recorded as $K_1$) is obtained:

$$K_1 = \frac{1}{2}\left(R_1 - 2R_2 \pm \sqrt{(R_1 - 2R_2)^2 - 4(R_2^2 + R_3^2)}\right). \tag{5}$$

However, the symbol '$\pm$' cannot be determined. $K$ is inherently dependent on the object that remains to be reconstructed. Therefore, Eq. (5) cannot be used directly as the solution of $K$. On our technical settings, the small window $w$ is located in the middle (or near middle) of the focal plane, allowing the pass of only the object's spectrum with low spatial frequency and blocking the high spatial frequency part that lies outside the window. Therefore, it is predictable that the generated amplitude $K$ is very smooth, and the phase $P$ is very flat (small aberration). This inference reveals that Eq. (5) is not suitable for solving $K$, because $K_1$ given by Eq. (5) is surely unsmooth, contrary to the fact that $K$ has a smooth shape. Moreover, the signal inside the window comes from the central Airy spot pattern, which has a smooth shape and tiny phase distortion. This contribution reinforces the results that $K$ is very smooth and $P$ is very flat (even close to 0). Based on this inference, the shape of $K$ can be approximated by the following formula, recorded as $K_2$, i.e.,

$$K = \left|\mathsf{F}\left(he^{i\alpha}w\right)\right| \propto \left|\mathsf{F}\left(w\right)\right| = K_2. \tag{6}$$

Obviously, $K_1$ is exact in the scale though it is indefinite, while $K_2$ is very precise in shape. This study takes a combined use of $K_1$ and $K_2$ to obtain a new approximation of $K$ with higher accuracy, $K \approx K_2 \cdot K_1^{max}/K_2^{max}$, where the superscript '*max*' means the maximum. In our previous research [26], it is found *K* does not need to be solved with high accuracy, because it has few effects on the final phase recovery, especially when there is subsequent iteration optimization. For example, we could even consider it as 0 to simplify the solution process in practical experiments. Now that *K* has been well approached and *P* is intrinsically flat and is considered as 0, an algebraic solution to the object phase *E* is achieved through Eq. (4), as expressed in Table 1 Section I. With this method, an accurate recovery of the object's amplitude and an approximate recovery of the object's phase are obtained.

The above is the theoretical establishment of the algebraic phase-retrieval solution. The core of this method is phase modulation on the object's spectrum within a small window. The window should be sufficiently small to ensure that the generated reference phase *P* is close to 0 and hence becomes known. A smaller window aids in achieving a more precise phase imaging result. Typically, an area located within the object's Airy spot can serve as a good phase shifting window. The validity of the algebraic phase-retrieval solution is verified by an experiment example of detecting rough phase objects. Sometimes there is non-negligible phase aberration on the background illumination which will be superposed into the final object's phase recovery. A practical solution for achieving a more accurate phase detection in the existence of background error is to test the background using the same technique before the experiment, and then remove the solved background by dividing by the amplitude and subtracting the phase. This approach is proved feasible by another experiment example of detecting a vortex phase object.

### *2.3 Accurate phase imaging*

Taking the approximate phase recovery solved by algebraic phase retrieval as the initial input, apply a powerful iterative optimization strategy to improve the phase recovery further. It is found the iteration shows a fast and stable convergence. Moreover, it exhibits a linear convergence (first-order convergence).

Therefore, we can always obtain an exact phase recovery provided the iteration is enough. In general, ten iterations are adequate to get a highly accurate phase recovery.

The whole flowchart of the non-interferometric high-accuracy QPI method is described in Table 1 below. It involves two sections. The first section is the acquisition of a good approximate phase recovery, which is obtained already by the algebraic phase retrieval method. The second section is the improvement to that approximate solution by a high-efficiency iterative algorithm. The algorithm is designed based on the alternative projection [38]. By propagating the optical field forward and backward ceaselessly and replacing the calculated intensity with the measured one, the phase becomes accurate gradually. Note that the directly-updated phase in the iteration is the reference phase $P$ rather than the object phase $E$. The object phase $E$ is calculated from $P$ based on the established exact intensity-phase relationship as shown in Eq. (4). This is a significant difference of the developed iteration algorithm from the famous Gerchberg-Saxton (GS) algorithm. Since the reference phase $P$ is much more smooth than the object phase $E$, the developed iteration algorithm that updates $P$ converges at a much higher speed than the GS algorithm that updates $E$ directly. The convergence speed of the iterative algorithm is represented by $p$ and $r$, which have a definition as

$$\lim_{k \to \infty} \left( \frac{\varepsilon_{k+1}}{\varepsilon_k^p} \right) = r. \tag{7}$$

In this equation, $\varepsilon_k$ and $\varepsilon_{k+1}$ are the root-mean-squared errors (RMS) of the $k^{th}$ and $(k+1)^{th}$ phase recovery, respectively. $p$ is a constant that defines the convergence order, and $r$ is another constant that defines the convergence rate. The phase RMS is defined as follows, where $E_R$ and $E$ are the recovered and inputted object phase, respectively, and $N$ is the number of all pixels.

$$\varepsilon = \sqrt{\sum (E_R - E)^2 / N}. \tag{8}$$

Table 1. Algorithm of non-interferometric accurate QPI for general complex objects

**Input -** three phase-shifts and the measured intensities $w$, $\{t_1, t_2, t_3\}$, $\{I_1, I_2, I_3\}$

**Condition** - available for arbitrary complex-valued objects, $w$ must be small in size

**Section I - an approximate solution obtained by algebraic phase retrieval**

1) Inverse linear transform: $\begin{bmatrix} R_1 \\ R_2 \\ R_3 \end{bmatrix} = \begin{pmatrix} 1 & 2\cos t_1 - 2 & 2\sin t_1 \\ 1 & 2\cos t_2 - 2 & 2\sin t_2 \\ 1 & 2\cos t_3 - 2 & 2\sin t_3 \end{pmatrix}^{-1} \begin{bmatrix} I_1 \\ I_2 \\ I_3 \end{bmatrix}$

2) Approximation of $Ke^{iP}$:

$$\begin{cases} K = |\mathscr{F}(w)| \dfrac{\left[ R_1 - 2R_2 + \sqrt{(R_1 - 2R_2)^2 - 4(R_2^2 + R_3^2)} \right]_{\max}}{2|\mathscr{F}(w)|_{\max}} \\ P = 0 \end{cases} \quad \text{or} \quad \begin{cases} K = 0 \\ P = 0 \end{cases}$$

3) Algebraic phase recovery: $\overline{C} = \sqrt{R_1}, \quad E_1 = P + \arctan\left[ R_3 / (R_2 + K^2) \right]$

**Output** - an accurate amplitude recovery $\{\overline{C}\}$ and an approximate phase recovery $\{E_1\}$

**Section II - iterative optimization of the object phase** (efficiency: linear convergence)

**Initialization** $E_2 = E_1$

**General Step**

1) Update $K$ and $P$ meanwhile: $Ke^{iP} = \mathscr{F}\left[ \mathscr{F}\left(\overline{C} e^{iE_2}\right) \cdot w \right]$

2) New phase recovery: $E_2 = P + \arctan\left[ R_3 / (R_2 + K^2) \right]$

**Executing** steps (1~2) in a loop for several times

**Output** - an accurate phase recovery $\{E_2\}$

In the iterative algorithm, both $K$ and $P$ are updated with the iteration simultaneously. This is a dual-drive optimization. As $K$ is being updated continuously, the phase difference $(E - P)$ solved by $\arctan[R_3 / (R_2 + K^2)]$ is increasingly precise. And as the reference phase $P$ is being improved meanwhile, the object's phase recovery $E_R$ becomes accurate quickly. Note that the initial input of $K$ can be arbitrary: either a reasonable approximation given by Eq. (5, 6) or 0. It affects only the iteration progress but does not influence the convergence speed. Because of the linear convergence property, the final phase recovery is highly accurate.

## 3. Simulation

### *3.1 Recovery of a complex object*

For a brief presentation, it is assumed the illumination has a uniform distribution: $B \equiv 1$. The complex-valued object is designed having large aberrations on both amplitude and phase, as shown in Fig. 2. The phase has a large range as $E \in [0, 4]$ rad. The minimum amplitude is set to 0.01 instead of 0 to ensure the phase has a meaningful value. In this simulation, the maximum amplitude is 100 times of its minimum. The simulation is performed on the platform of MATLAB 2015b. All the data are discretized with a size of 256×256. The 2D Fourier transform is applied to generate the object's Fourier spectrum and the output intensities based on Eq. (1). In this calculation, the inputs $\{B, C, E\}$ are extended into 1536×1536 by zero-padding to obtain a higher spatial frequency resolution. The output intensity $I$ is produced by shifting the phase of the object's Fourier spectrum within a small window area $w$. As shown in Fig. 2(c), this is a square area with the size of 7×7 pixels that occupies only $2 \times 10^{-5}$ the area of whole but carries more than 21% the intensity.

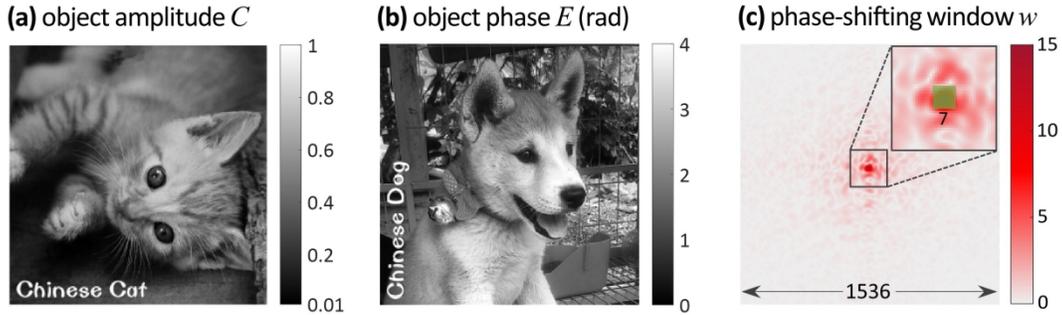

Fig. 2. The tested complex-valued object and the phase-shifting window used in the numerical simulation. (a) object amplitude. (b) object phase with phase range [0, 4] rad. (c) The phase-shifting window, which locates on the centre of the Fourier spectrum, occupying 7×7 pixels among 1536×1536 pixels of the whole region.

The phase-shifting values are set as $t = \{0, \pi/2, \pi\}$, and the corresponding output intensities are simulated applying Eq. (1) and $I = U^2$. As shown in the left-hand figure of Fig. 3(a), the output amplitude $\sqrt{I(t)}$ under $t = 0$ is exactly the object amplitude, where the phase information is hidden completely. When there are phase shifts $t = \{\pi/2, \pi\}$ applied to the object's Fourier spectrum, the phase emerges gradually in the output image but suffers severe interference by the amplitude. Then according to the algebraic phase-retrieval solution described in Table 1 Section I, the object phase is reconstructed successfully, as shown in Fig. 3(b).

Though it appears a good phase recovery, there is still significant residual error seen from Fig. 3(c). This error can be reduced continuously by the developed iterative optimization expressed in Table 1 Section II. As depicted in Fig. 3(d), after applying 25 iterations, the new phase recovery appears almost the same with the inputted one. Quantitatively, the phase RMS value drops largely from 0.2103 to 0.0003, about 1‰ of the original. This significant improvement on the phase recovery accuracy after iterative optimization is also seen from the negligible residual error shown in the right-hand figure of Fig. 3(d).

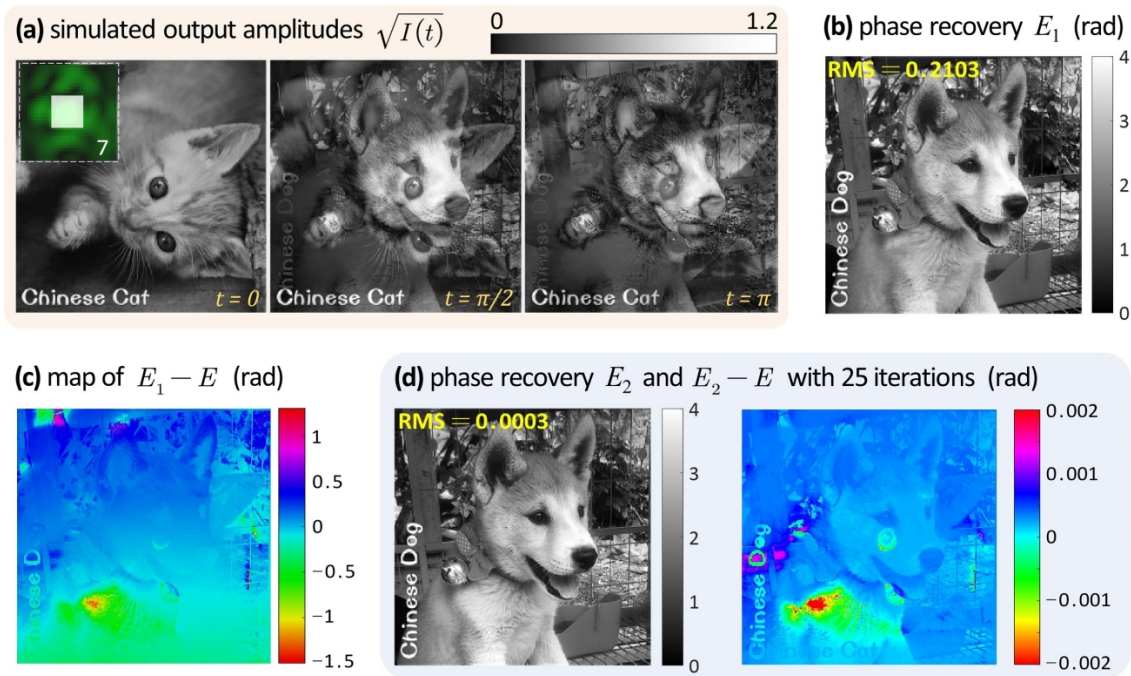

Fig. 3. The performance of the proposed non-interferometric high-accuracy QPI method in recovering a complex-valued object. (a) the measured output intensity patterns under three different phase-shifting values. (b) the algebraic phase recovery. (c) residual error of (b). (d) the improved phase recovery obtained after applying 25 iterations.

## *3.2 Linear-convergence property*

The convergence performance of the iteration is illustrated in the left-hand figure of Fig. 4. It is seen the RMS value shown in the logarithmic scale goes down with a nearly constant speed. According to the definition of Eq. (8), the convergence order $p$ is determined exactly as $p = 1$. Due to this linear convergence property, it is anticipated that the phase recovery error will decrease until it approaches zero unlimitedly.

Therefore, the exact phase recovery is achievable with sufficient iterations. According to the phase RMS ratio curve, it is evaluated the convergence rate $r$ is around 0.778. Further, a series of simulations on diverse complex-valued objects were performed and have proved that the linear convergence is an intrinsic property of Table 1 Section II. First, the convergence speed on different phase ranges is studied. This simulation takes the same settings as given in Fig. 2 but resets the phase range. As shown in the disk-marked curve depicted in the right-hand figure of Fig. 4, all the iterations show linear convergence property but have different convergence rates. A basic conclusion is that the convergence becomes slower with the increase of the phase range. This could be explained in the view of distribution change of the object's Fourier spectrum. A complex-valued object with a larger phase range tends to have a more dispersed and speckled spectrum. Hence, the reference phase $P$ will not be so much close to 0, which leads to a less precise phase recovery ultimately. Second, the convergence speed for pure-phase objects is investigated. The inputted object phase is the same as that in Fig. 2, but the object amplitude now is $C \equiv 1$ instead of a cat image. As shown in the square-marked curve of Fig. 4, the iteration also shows a linear convergence, and even the convergence rate follows a quite similar trace with that of complex-valued objects. Third, the effect the phase-shifting window size has on the convergence rate is investigated. Several window widths, $W_d = \{3, 5, 7, 9\}$ pixels, were taken into consideration. As seen in the right-hand figure of Fig. 4, all the iterations exhibit linear convergence property. The convergence rate under different $W_d$ appears a very similar tendency with each other with the increase of the object phase range. The smaller the window size is, the faster the iteration converges. This could be explained as follows. As shown in the inset figure of the right-hand figure of Fig. 4, the object's spectrum inside a smaller window appears flatter than that of a larger window. So the reference phase $P$ produced by a smaller window is smoother and thus yields a more precise phase recovery. However, a too small phase-shifting window makes the output intensities indistinguishable so that increases the measurement difficulty. Therefore, there is a trade-off between the phase recovery accuracy and the experimental feasibility. Generally, it is recommended that the phase-shifting window *w* occupies about half area of the Airy spot.

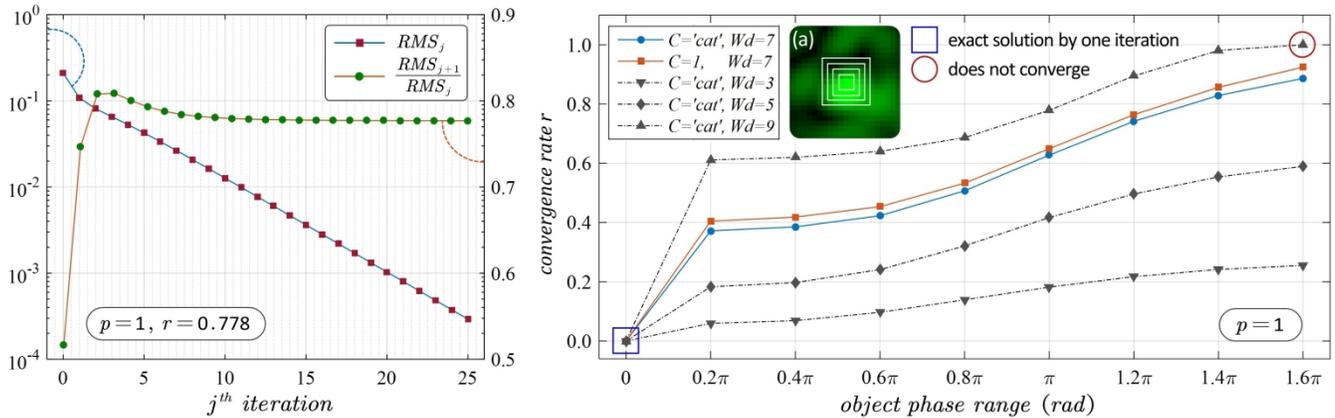

Fig. 4. The convergence performance of the iterative optimization strategy. The left figure shows the variations of phase RMS and phase RMS ratio along with the iteration. The right figure shows the convergence rate *r* of the linear convergence (*p* = 1) under diverse simulation settings. In this figure, *C* = *cat* and *C* = *1* means a complex-valued object taking the cat image shown in Fig. 2(a) as the amplitude and a pure-phase object, respectively. Due to the inaccurate RMS calculation caused by phase wrapping, simulation with phase range over $1.6\pi$ were not investigated.

In summary, the linear convergence shown here is an intrinsic characteristic of the developed iterative optimization strategy Table 1 Section II. This property ensures us obtain the exact phase recovery. To accelerate the convergence, a small enough phase-shifting window is required. Otherwise, the iteration may fall stagnant or even be divergent instead of being a linear convergence.

## 4. Experiments

### *4.1 Experimental platform*

The experimental optical setup is arranged as illustrated in Fig. 1. Both the unknown phase object and the small-window phase modulation are generated by reflective phase-only SLMs (*UPOLabs, HDSLM63R, 1280×720, 8 bits, 6.37-μm pixel width*). SLM #1 is placed on the object plane to load the phase object and SLM #2 at the focal plane to load the phase mask. The incident beam is emitted from a commercial single-mode laser (*Changchun New Industries, MGL-III-532, 532±1 nm, TEM$_{00}$ mode*). A pinhole with 10-μm diameter is placed behind to filter the original laser beam to be smooth and aberration-free. To match the SLM screen, the parallel beam is truncated by a 3mm-diameter pupil. A polarizer is added before the SLM to ensure the SLM operating in phase-only mode. On the focal plane, the object's Fourier spectrum is shifted in phase by SLM #2 within a small window area. Subsequently, a CMOS image sensor (*Microview, HK-A5100-GM17, 2560×2160, 16 bit, 6.5-μm pixel width*) is applied to record the output intensity pattern. In this experiment, three phase-shifting values $t = \{0, \pi/2, \pi\}$ are loaded on SLM #2 successively, and correspondingly, three intensity images $\{I_1, I_2, I_3\}$ are measured by the CMOS. Then the phase recovery performance of the proposed general algebraic phase retrieval method (expressed in Table 1 Section I) is investigated and evaluated. There is always an unmodulated light beam that merges into the modulated beam due to the pixel gap of the SLM and hence causes a blend. To eliminate aliasing, an additional suitable oblique phase mask is added into both SLM #1 and SLM #2 to separate the modulated light beam into the measurement.

## *4.2 Rough phase object test*

At first, we test the performance of the method in detecting rough phase objects, and meanwhile, the spatial resolution is investigated. Theoretically, from the given beam width ($D$ = 3 mm), imaging distance ($f$ = 150 mm), and the wavelength ($\lambda$ = 532 nm), the spatial resolution of current optical imaging platform is determined as $d = \lambda f / D = 26.65$ μm. As shown in Fig. 5(a), a grayscale image 'balls' is inputted to SLM #1 as the rough phase object with a phase range of [0, 2π] rad. The phase modulation is set as $t = \{0, \pi/2, \pi\}$. The recorded intensity images are listed in Fig. 5(b), shown in log-scale to present a clearer distinction. Due to the system error and experimental defects including the SLM not operating on phase-only mode, a limited numerical aperture and a poor 4*f*-imaging-system configuration, the phase information emerges when there is no phase modulation ($t$ = 0). However, that cannot provide us quantitative phase detection. With the help of two additional intensity images, as shown in Fig. 5(c), the phase object is well recovered taking only one algebraic calculation. It is a high-accuracy reconstruction and exhibits many details. As shown in Fig. 5(d), from the comparison between the inputted phase and the recovered phase along the section lines given by Fig. 5(a), it is estimated that the phase recovery error in average is about 0.1 rad, and the spatial resolution is better than 40 μm, a value matches the theoretical expectation.

In addition, there are two noteworthy phenomena in this experiment. First, the phase recovery solution is point-to-point, so it is workable for selecting any area of the intensity data into the calculation. For example, here the central data of 501×501 pixels are selected from the whole region of 736×736 pixels for the object reconstruction, and that works well. Second, the phase recovery can avoid suffering the influence of the meaningless intensity data caused by the dead pixels. For instance, there is a 'branch' marked by an arrow in Fig. 2(b), but it is missing in the recovered phase in Fig. 2(c), showing a self-healing ability of the phase.

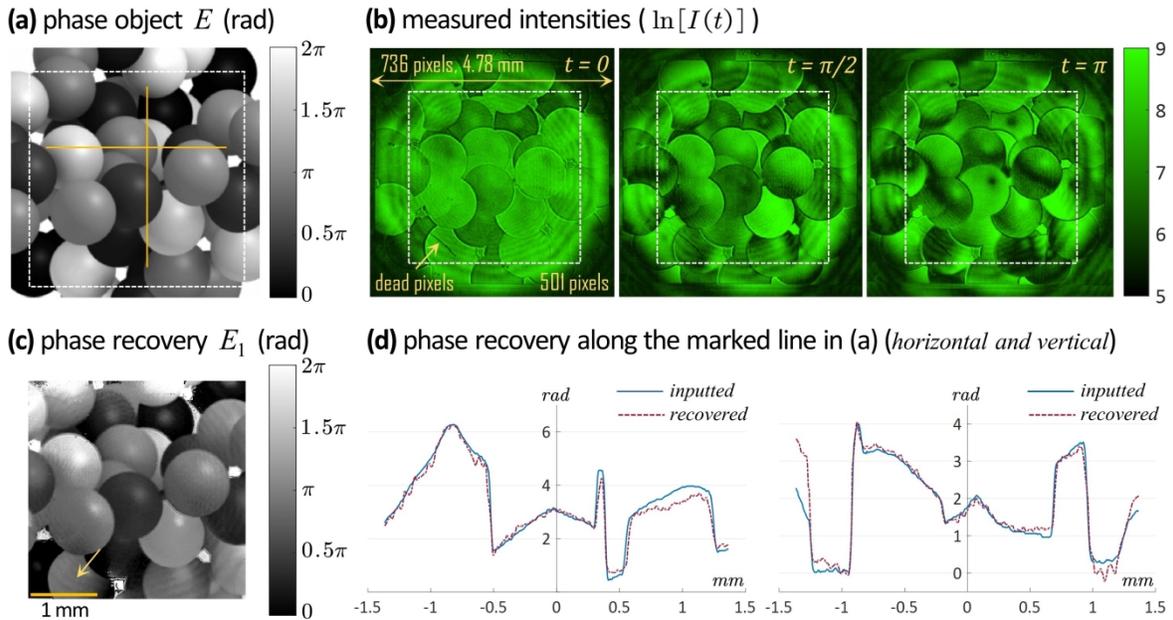

Fig. 5. Experimental phase imaging results of a rough phase object. (a) inputted phase on SLM #1, a bunch of balls with a phase range of [0, 2π] rad. (b) recorded intensity patterns by the CMOS. (c) algebraic phase recovery. (d) comparison between the inputted phase and the recovered phase.

### 4.3 Vortex phase object test

In optical communication based on orbital-angular-momentum (OAM), it is important to reconstruct the phase distribution carried by the vortex beam so as to decode the digital information [40, 41]. The proposed method is able to detect the phase information of a vortex beam with high efficiency. Here we arrange a phase imaging test to a vortex beam. In addition to determining topological number of the vortex phase, meanwhile the accuracy improvement brought by background elimination is also investigated. As seen in Fig. 5(b, c), there are some obvious ring artifacts caused by near-field diffraction in both the recorded intensity patterns and the recovered phase image, which reduces the phase recovery accuracy. One effective means to remove this kind of artifact is the background elimination strategy. To conduct this strategy, two phase objects are loaded on SLM #1. One is the unknown vortex phase object to be reconstructed, and the other is a zero-phase that is used to detect the background amplitude and phase aberration.

As listed in Fig. 6, the experimental results are shown in detail. At first, six intensity patterns are measured, as given by Fig. 6(b, e), where the phase modulation is set as $t = \{0, \pi/2, \pi\}$. As seen in the algorithm given by Table 1, the two inputted phase objects are reconstructed from their intensity patterns, respectively, as shown in Fig. 6(c) and Fig. 6(f). For the vortex phase, it is seen the reconstructed phase appears obvious vortex structure and has a topological number of 16. It exhibits the same vortex structure and even a very similar phase shape compared to the inputted one. However, there are obvious phase errors on the edge. Besides, there are many ring artifacts emerging on the measured intensities and the recovered phase. By inputting a zero-phase object, the background amplitude and phase of the laser beam are detected. It is seen in Fig. 6(f) that the background phase aberration appears an oblique-plane shape with a small range of variation. After removing the background, a more accurate phase recovery is obtained, as shown in Fig. 6(g, h). Now the ring artifacts caused by Fresnel diffraction are basically eliminated. Fig. 6(g) indicates the laser beam distribution clearly, and it shows the SLM has a nearly homogenous modulation on the light intensity, but blocks the light significantly at the vortex angles. Fig. 6(h) demonstrates a good phase recovery that is close to the inputted phase. The original phase recovery errors on the edge are eliminated significantly, resulting in a improved phase recovery. The optimization of phase recovery accuracy is quantified by Fig. 6(i), where the phase RMS drops from 0.81 to 0.39 after background elimination. Similar to the experiment shown in Fig. 5, the small speckles on the CMOS screen affected only the intensity but has little influence on the recovered phase. Overall, this experiment verifies the potential of the proposed method in detecting vortex phases. Meanwhile, the accuracy improvement brought by the background elimination strategy is also validated. Note that such background calibration needs to be done only once before the experiment, so little extra work is needed.

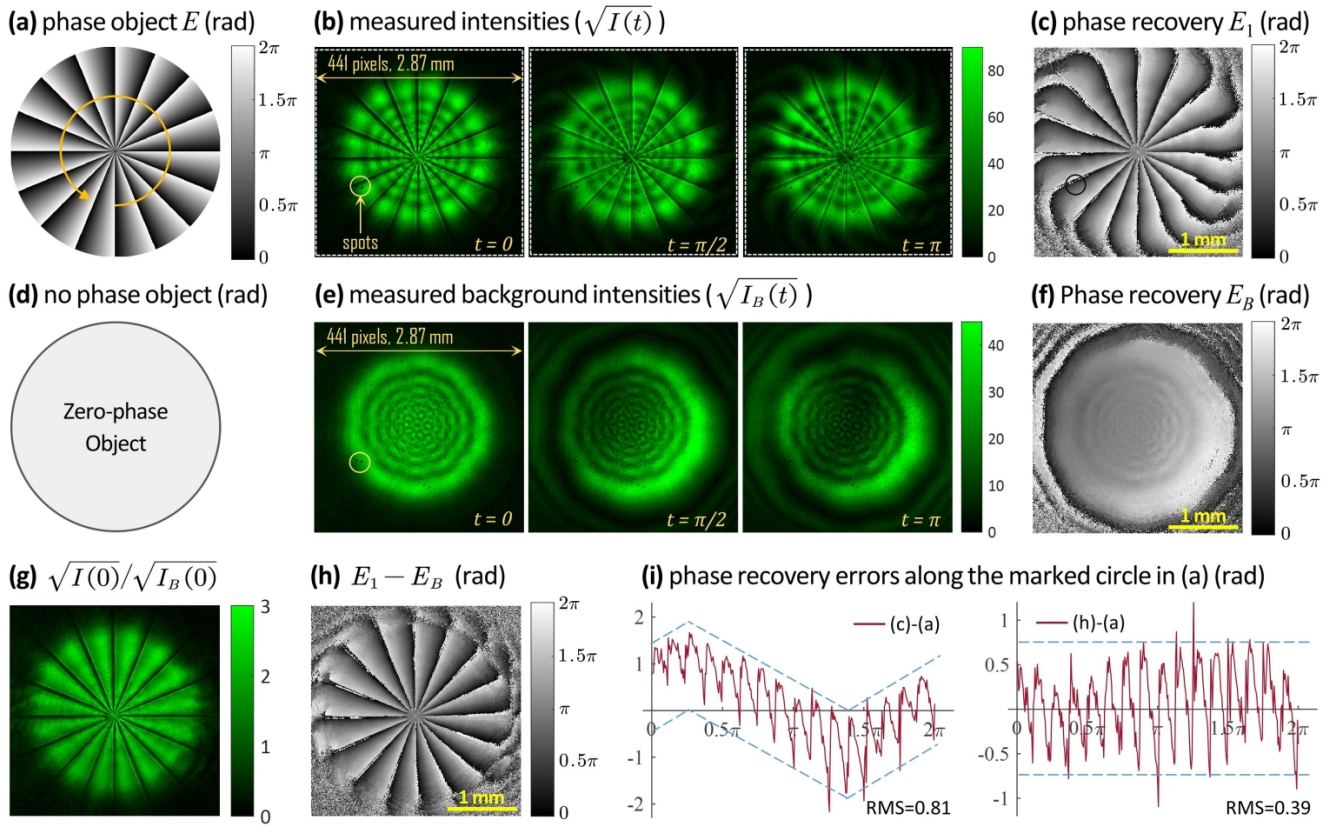

Fig. 6. Experimental phase imaging results of a vortex phase object. (a) inputted phase on SLM #1, a vortex phase plate with a topological number of 16. (b) recorded intensity patterns by the CMOS. (c) algebraic phase recovery. (d) inputted zeros-phase object. (e) recorded background intensity patterns. (f) algebraic phase recovery of the background. (g) new amplitude recovery. (h) new phase recovery. (i) phase recovery errors of (c) and (h), where the phase RMS drops from 0.81 to 0.39 after background elimination.

## 5. Conclusion

A general non-interferometric accurate QPI technique is proposed, which has much potential to be applied in non-interferometric phase imaging. The optical configuration is similar to a typical 4*f* phase contrast platform, to which we have made significant modifications on phase-shifting area, phase-shifting value, and phase retrieval algorithm. By combining algebraic phase retrieval and a high-efficiency iterative optimization, the proposed method is capable of achieving an accurate object reconstruction for arbitrary complex-objects with three intensity measurements. Based on a rigorous algebraic relationship built between object phase and the measured intensities, a good approximate phase recovery is obtained. Afterwards, an efficient iterative optimization algorithm is developed to improve that approximate phase recovery. Due to the linear-convergence property of the iteration, the error caused by regarding the reference phase as zero could be eliminated completely, and finally, an exact phase recovery is achieved. This method is a significant progress of previous works, including generalized phase contrast [28] and phase-retrieval algebraic solution [26]. The combination of algebraic phase retrieval and iterative optimization is previously studied in [42], where the phase recovery accuracy gets significant improvement with the help of the angle relationship established by algebraic phase retrieval. In contrast, the study shows the approximate phase recovery obtained by algebraic phase retrieval can be optimized into an accurate one quickly through iterative optimization, which is also actually a result of the excellent initial phase guess provided by the algebraic phase retrieval in turn. In brief, this study develops a high-accuracy QPI technique for non-interferometric quantitative phase imaging, and the future work will focus on extending the method to 3D application cases.

## 6. Funding, acknowledgments, and disclosures

### *6.1 Funding*

National Science Foundation of China (NSFC), U1931137; China Scholarship Council (CSC) during a visit of Jianhui Huang to Georgia Tech is also acknowledged.

### *6.2 Acknowledgments*

The authors sincerely appreciate the help of UPOLabs (http://www.realic.cn) on the optical experimental setup and test.

*6.3 Disclosures*

The authors declare no conflicts of interest.
# References

1. K. G. Phillips, S. L. Jacques, and O. J. T. McCarty, "Measurement of Single Cell Refractive Index, Dry Mass, Volume, and Density Using a Transillumination Microscope," Physical Review Letters **109**, 118105 (2012).

2. B. Bhaduri, C. Edwards, H. Pham, R. Zhou, T. H. Nguyen, L. L. Goddard, and G. Popescu, "Diffraction phase microscopy: principles and applications in materials and life sciences," Advances in Optics and Photonics **6**, 57-119 (2014).

3. Y. Park, C. Depeursinge, and G. Popescu, "Quantitative phase imaging in biomedicine," Nature Photonics **12**, 578-589 (2018).

4. A. Pan, C. Zuo, and B. Yao, "High-resolution and large field-of-view Fourier ptychographic microscopy and its applications in biomedicine," Reports on Progress in Physics **83**, 096101 (2020).

5. S. K. Debnath and Y. Park, "Real-time quantitative phase imaging with a spatial phase-shifting algorithm," Optics Letters **36**, 4677-4679 (2011).

6. M. H. Jenkins and T. K. Gaylord, "Three-dimensional quantitative phase imaging via tomographic deconvolution phase microscopy," Applied Optics **54**, 9213-9227 (2015).

7. Y. Bao and T. K. Gaylord, "Iterative optimization in tomographic deconvolution phase microscopy," Journal of the Optical Society of America A **35**, 652-660 (2018).

8. J. M. Soto, J. A. Rodrigo, and T. Alieva, "Label-free quantitative 3D tomographic imaging for partially coherent light microscopy," Optics Express **25**, 15699-15712 (2017).

9. J. Li, Q. Chen, J. Sun, J. Zhang, J. Ding, and C. Zuo, "Three-dimensional tomographic microscopy technique with multi-frequency combination with partially coherent illuminations," Biomedical Optics Express **9**, 2526-2542 (2018).

10. M. Chen, D. Ren, H.-Y. Liu, S. Chowdhury, and L. Waller, "Multi-layer Born multiple-scattering model for 3D phase microscopy," Optica **7**, 394-403 (2020).

11. P. Marquet, C. Depeursinge, and P. J. Magistretti, "Exploring Neural Cell Dynamics with Digital Holographic Microscopy," Annual Review of Biomedical Engineering **15**, 407-431 (2013).

12. T. Ikeda, G. Popescu, R. R. Dasari, and M. S. Feld, "Hilbert phase microscopy for investigating fast dynamics in transparent systems," Optics Letters **30**, 1165-1167 (2005).

13. W. Xu, M. H. Jericho, I. A. Meinertzhagen, and H. J. Kreuzer, "Digital in-line holography of microspheres," Applied Optics **41**, 5367-5375 (2002).

14. W. Luo, A. Greenbaum, Y. Zhang, and A. Ozcan, "Synthetic aperture-based on-chip microscopy," Light: Science & Applications **4**, e261-e261 (2015).

15. J. Zhang, J. Sun, Q. Chen, J. Li, and C. Zuo, "Adaptive pixel-super-resolved lensfree in-line digital holography for wide-field on-chip microscopy," Scientific Reports **7**, 11777 (2017).

16. T. Kakue, R. Yonesaka, T. Tahara, Y. Awatsuji, K. Nishio, S. Ura, T. Kubota, and O. Matoba, "High-speed phase imaging by parallel phase-shifting digital holography," Optics Letters **36**, 4131-4133 (2011).

17. S. H. S. Yaghoubi, S. Ebrahimi, M. Dashtdar, A. Doblas, and B. Javidi, "Common-path, single-shot phase-shifting digital holographic microscopy using a Ronchi ruling," Applied Physics Letters **114**, 183701 (2019).
20

18. N. Patel, V. Trivedi, S. Mahajan, V. Chhaniwal, C. Fournier, S. Lee, B. Javidi, and A. Anand, "Wavefront division digital holographic microscopy," Biomedical Optics Express **9**, 2779-2784 (2018).
19. S. Ebrahimi, M. Dashtdar, A. Anand, and B. Javidi, "Common-path lensless digital holographic microscope employing a Fresnel biprism," Optics and Lasers in Engineering **128**, 106014 (2020).
20. G. Popescu, T. Ikeda, R. R. Dasari, and M. S. Feld, "Diffraction phase microscopy for quantifying cell structure and dynamics," Optics Letters **31**, 775-777 (2006).
21. S. Ajithaprasad and R. Gannavarpu, "Non-invasive precision metrology using diffraction phase microscopy and space-frequency method," Optics and Lasers in Engineering **109**, 17-22 (2018).
22. P. Gao, I. Harder, V. Nercissian, K. Mantel, and B. Yao, "Phase-shifting point-diffraction interferometry with common-path and in-line configuration for microscopy," Optics Letters **35**, 712-714 (2010).
23. R. Park, D. W. Kim, and H. H. Barrett, "Synthetic phase-shifting for optical testing: Point-diffraction interferometry without null optics or phase shifters," Optics Express **21**, 26398-26417 (2013).
24. D. Palima and J. Glückstad, "Generalised phase contrast: microscopy, manipulation and more," Contemporary Physics **51**, 249-265 (2010).
25. C. Zuo, Q. Chen, W. Qu, and A. Asundi, "Noninterferometric single-shot quantitative phase microscopy," Optics Letters **38**, 3538-3541 (2013).
26. J. Huang, H. Jin, Q. Ye, and G. Meng, "Phase-Retrieval Algebraic Solution Based on Window Modulation," Annalen der Physik **530**, 1800063 (2018).
27. F. Zernike, "Phase contrast, a new method for the microscopic observation of transparent objects part II," Physica **9**, 974-986 (1942).
28. P. J. Rodrigo, D. Palima, and J. Glückstad, "Accurate quantitative phase imaging using generalized phase contrast," Optics Express **16**, 2740-2751 (2008).
29. P. Gao, B. Yao, I. Harder, N. Lindlein, and F. J. Torcal-Milla, "Phase-shifting Zernike phase contrast microscopy for quantitative phase measurement," Optics Letters **36**, 4305-4307 (2011).
30. J. K. Wallace, S. Rao, R. Jensen-Clem, and G. Serabyn, *Phase-shifting Zernike interferometer wavefront sensor*, SPIE Optical Engineering + Applications (SPIE, 2011), Vol. 8126.
31. Z. Wang, L. Millet, M. Mir, H. Ding, S. Unarunotai, J. Rogers, M. U. Gillette, and G. Popescu, "Spatial light interference microscopy (SLIM)," Optics Express **19**, 1016-1026 (2011).
32. N. Streibl, "Phase imaging by the transport equation of intensity," Optics Communications **49**, 6-10 (1984).
33. C. Zuo, Q. Chen, Y. Yu, and A. Asundi, "Transport-of-intensity phase imaging using Savitzky-Golay differentiation filter - theory and applications," Optics Express **21**, 5346-5362 (2013).
34. C. Wang, Q. Fu, X. Dun, and W. Heidrich, "Quantitative Phase and Intensity Microscopy Using Snapshot White Light Wavefront Sensing," Scientific Reports **9**, 13795 (2019).
35. P. Berto, H. Rigneault, and M. Guillon, "Wavefront sensing with a thin diffuser," Optics Letters **42**, 5117-5120 (2017).
36. L. Lu, J. Sun, J. Zhang, Y. Fan, Q. Chen, and C. Zuo, "Quantitative Phase Imaging Camera With a Weak Diffuser," Frontiers in Physics **7**(2019).
37. C. Zuo, Q. Chen, W. Qu, and A. Asundi, "High-speed transport-of-intensity phase microscopy with an electrically tunable lens," Optics Express **21**, 24060-24075 (2013).
38. Y. Shechtman, Y. C. Eldar, O. Cohen, H. N. Chapman, J. Miao, and M. Segev, "Phase Retrieval with Application to Optical Imaging: A contemporary overview," IEEE Signal Processing Magazine **32**, 87-109 (2015).
39. A. G. Robert, "Phase Retrieval And Diversity In Adaptive Optics," Optical Engineering **21**, 829-832 (1982).